\documentclass[fleqn,twoside]{article}
\usepackage{espcrc2}
\usepackage{axodraw}
\usepackage{epsfig}
\readRCS
$Id: espcrc2.tex,v 1.2 2004/02/24 11:22:11 spepping Exp $
\usepackage{graphicx}
\usepackage[figuresright]{rotating}

\newcommand{\AmS}{{\protect\the\textfont2
  A\kern-.1667em\lower.5ex\hbox{M}\kern-.125emS}}

\title{Evaluating multiloop Feynman integrals by
Mellin-Barnes representation}

\author{V.A.~Smirnov\address[MCSD]{
Nuclear Physics Institute of Moscow State University,\\
Moscow 119992, Russia}%
        \thanks{Supported by
DFG Mercator visiting professorship No.
Ha 202/110-1 and Volkswagen Foundation,
Contract No. I/77788.}}

\newcommand{\be}{\begin{equation}}
\newcommand{\ee}{\end{equation}}
\newcommand{\bea}{\begin{eqnarray}}
\newcommand{\eea}{\end{eqnarray}}

\newcommand{\Gm}{\Gamma}

\newcommand{\ep}{\epsilon}

\newcommand{\lm}{\lambda}

\newcommand{\dd}{\mbox{d}}

\newcommand{\nn}{\nonumber}

\begin{document}

\begin{abstract}
The status of analytical evaluation of double and triple
box diagrams is characterized.
The method of Mellin-Barnes representation as a tool to evaluate
master integrals in these problems is advocated. New
MB representations for massive on-shell double boxes
with general powers of propagators are presented.
\end{abstract}

\maketitle

\section{Introduction}

When calculating physical quantities that describe a given
process one needs to evaluate a lot of Feynman integrals.
After a tensor reduction based on some projectors (see, e.g.,
\cite{tenred}) a given Feynman graph generates
various scalar Feynman
integrals that have the same structure of the integrand with
various distributions of powers of propagators.
A straightforward analytical strategy is
to evaluate, by some methods, every scalar Feynman
integral generated by the given graph.
If the number of these integrals is small such strategy is
reasonable.
In non-trivial situations, where the number of different
integrals can be at the level of hundreds and thousands,
it is reasonable to follow a well-known advanced strategy:
to derive, without calculation,
and then apply integration by parts (IBP) \cite{IBP}
and Lorentz-invariance (LI) \cite{LI}
identities between the given family of Feynman integrals as
{\it recurrence relations}.
The goal of this procedure is to express a general integral from the
given family as a linear combination of some basic
({\it master}) integrals.
Therefore the whole problem of evaluation is decomposed into two
parts:  solution of the reduction procedure and
evaluation of the master Feynman integrals.

There were several recent attempts to make the reduction
procedure systematic:

({\em i}) Using the fact that
the total number of IBP and LI equations grows faster
than the number of independent Feynman integrals one can
sooner or later obtain an overconstrained system
of equations \cite{LGR1,LGR2}.

({\em ii}) Using relations that can be obtained by
tricks with shifting dimension \cite{Tar}.

({\em iii}) Baikov's method \cite{Bai}.

({\em iv})
Another attempt in this direction is the use of Gr\"obner basis
(see, e.g., \cite{Groe}).

To evaluate master integrals one uses various methods.
In particular, for the evaluation of double and triple boxes
the following two methods were successfully applied last years:
the method of Mellin--Barnes (MB) representation \cite{MB}
and differential equations \cite{DE}.
The first of them is based on the following representation
\bea
\frac{1}{(X+Y)^{\lm}} = \frac{1}{\Gm(\lm)}
&&
\nn \\ &&  \hspace*{-27mm}
\times \frac{1}{2\pi  i }\int_{- i  \infty}^{+ i  \infty} \dd z
\frac{Y^z}{X^{\lm+z}} \Gm(\lm+z) \Gm(-z)
\label{MB}
\eea
which is applied to replace a sum of two terms raised
to some power by their products in some powers at the cost of
introducing an extra integration.
%The poles with a $\Gm(\ldots+z)$ dependence
%%(let us call them IR poles)
%are to the left of the contour and
%the poles with a $\Gm(\ldots-z)$ dependence %(UV poles)
%are to the right.
Usually one starts from alpha or Feynman parameters, then
introduces, in a minimal way, MB integrations, performs
internal integrations over Feynman parameters in terms of
gamma functions and obtains a multiple MB representation.
It is useful to derive such representation for general powers
of the propagators.

The standard procedure of taking residues and shifting contours is
used \cite{MB}, with the goal to obtain a sum of integrals where
one may expand integrands in Laurent series in $\ep=(4-d)/2$
(where $d$ is the space-time dimension within dimensional
regularization \cite{dimreg}). Then one
can use the first and the second Barnes lemmas and their
corollaries to perform some of the MB integrations explicitly.
In the last integrations which usually carry dependence
on the external variables, one closes contour in the complex plane
and sums up corresponding series.
(See \cite{S4} for details of the method).

A typical example of application of this method is the
evaluation of master integrals for massless on-shell
($p_i^2=0,\;i=1,2,3,4$) double
boxes \cite{MB,GTGR0ATT} where the reduction procedure was performed
using shifting dimension \cite{SV,AGORT}, with
multiple subsequent applications \cite{appl}.

\section{Triple boxes and double boxes with one leg off-shell}

\begin {figure} [htbp]
\begin{picture}(170,50)(-30,0)
\Line(-15,0)(0,0)
\Line(-15,50)(0,50)
\Line(165,0)(150,0)
\Line(150,0)(100,0)
\Line(150,50)(100,50)
\Line(165,50)(150,50)
\Line(0,0)(50,0)
\Line(50,0)(100,0)
\Line(100,50)(50,50)
\Line(50,50)(0,50)
\Line(0,50)(0,0)
\Line(50,0)(50,50)
\Line(100,0)(100,50)
\Line(150,0)(150,50)
%%%%%%%%%%%%%
\Vertex(0,0){1.5}
\Vertex(50,0){1.5}
\Vertex(100,0){1.5}
\Vertex(0,50){1.5}
\Vertex(50,50){1.5}
\Vertex(100,50){1.5}
\Vertex(150,0){1.5}
\Vertex(150,50){1.5}
%\Text(-22,0)[]{$p_1$}
%\Text(174,0)[]{$p_3$}
%\Text(-22,50)[]{$p_2$}
%\Text(174,50)[]{$p_4$}
%\Text(25,43)[]{\small 1}
%\Text(-7,25)[]{\small 2}
%Text(25,7)[]{\small 3}
%\Text(75,7)[]{\small 4}
%\Text(43,25)[]{\small 7}
%\Text(93,25)[]{\small 5}
%\Text(75,43)[]{\small 6}
%\Text(125,7)[]{\small 8}
%\Text(125,43)[]{\small 10}
%\Text(143,25)[]{\small 9}
\end{picture}
%\vspace*{2mm}\\
\label{3box}
\caption{Planar triple box diagram.}
\end{figure}
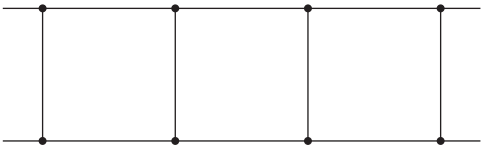

The first calculation of massless on-shell triple boxes in dimensional
regularization was done in \cite{3bR}, where
the Regge asymptotics (in the limit $t/s\to 0$) of the master planar
triple box shown in Fig.~\ref{3box}
was calculated with the help of the strategy of
expansion by regions \cite{BS}.
Later it became possible to evaluate it analytically \cite{3b},
with the help of a sevenfold MB representation,
in terms of harmonic polylogarithms \cite{HPL}.
In fact, any massless planar on-shell triple box can be evaluated
by this procedure so that evaluation of three-loop virtual
corrections to various scattering processes is not a miracle.

Let us mention that the necessity to analytically evaluate
massless on-shell double and triple boxes arose when
studying cross order relations in $N=4$ supersymmetric gauge
theories \cite{SUSY}. In fact, to check some factorization one needs
just one more triple box, in addition to  Fig.~1.

The complexity of the problem depends not only on the number of
loops but also on the number of parameters.
One way to include one more parameter into massless on-shell
double boxes is to consider one leg off shell, i.e.
$p_1^2=q^2\neq 0$, $p_i^2=0,\;i=2,3,4$.
This problem was successfully solved last years.
The reduction to master integrals was done using Laporta's idea
in \cite{LGR2}.
Master integrals were calculated using MB representation
(first results in \cite{S2})
and DE (systematic evaluation in \cite{LGR2}).
All results are expressed in terms of
two-dimensional harmonic polylogarithms  \cite{LGR2} which
generalize harmonic polylogarithms.
This combination of reduction based on \cite{LGR1,LGR2}
and DE was successfully applied in numerous calculations, e.g.,
various classes of vertex diagrams \cite{appl-vert}.

\section{Massive on-shell double boxes}

\begin {figure} [htbp]
\begin{picture}(170,50)(-30,0)
\Line(-15,0)(0,0)
\Line(-15,50)(0,50)
\Line(115,0)(100,0)
\Line(115,50)(100,50)
\Line(0,0)(50,0)
\Line(50,0)(100,0)
\Line(100,50)(50,50)
\Line(50,50)(0,50)
\DashLine(0,50)(0,0){3}
\DashLine(50,0)(50,50){3}
\DashLine(100,0)(100,50){3}
\Vertex(0,0){1.5}
\Vertex(50,0){1.5}
\Vertex(100,0){1.5}
\Vertex(0,50){1.5}
\Vertex(50,50){1.5}
\Vertex(100,50){1.5}
\Text(25,43)[]{\small 1}
\Text(-7,25)[]{\small 2}
\Text(25,7)[]{\small 3}
\Text(75,7)[]{\small 4}
\Text(43,25)[]{\small 7}
\Text(93,25)[]{\small 5}
\Text(75,43)[]{\small 6}
\end{picture}
\caption{Planar massive on-shell double box of the first type}
\label{2mboxPl1}
\end{figure}
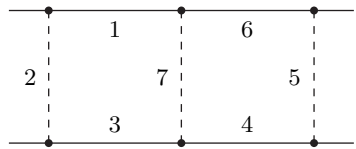

Another way to include one more parameter into massless on-shell
double boxes is to turn to massive on-shell double boxes with one
non-zero mass, $p_i^2=m^2,\;i=1,2,3,4$.
Keeping in mind Bhabha scattering let us distinguish
two planar and one non-planar diagrams as
three basic most complicated types of massive on-shell double boxes.
The first planar graph is shown in Fig.~\ref{2mboxPl1}.
For the corresponding general Feynman integral
\bea
B_{PL,1}(a_1,\ldots,a_8;s,t,m^2;\ep)
&& \nn \\ && \hspace*{-48mm}
=\int\int \frac{\dd^dk \, \dd^dl}{(k^2-m^2)^{a_1}
[(k+p_1)^2)]^{a_2}[(k-l)^2]^{a_7}
}
\nn \\ && \hspace*{-48mm}
\times
\frac{1}{[(k+p_1+p_2)^2-m^2]^{a_3}[(l+p_1+p_2+p_3)^2]^{a_5}}
\nn \\ && \hspace*{-48mm}
\times \frac{[(k+p_1+p_2+p_3)^2]^{-a_8}}{
[(l+p_1+p_2)^2-m^2)]^{a_4}
(l^2-m^2)^{a_6}}\,,
\label{2boxPL1}
\eea
with an irreducible numerator chosen as
$(k+p_1+p_2+p_3)^2$, the following
sixfold MB representation was derived in \cite{S3}:
\bea
B_{PL,1}(a_1,\ldots,a_8;s,t,m^2;\ep)
&& \nn \\ &&  \hspace*{-48mm}
= \frac{\left(i\pi^{d/2} \right)^2 (-1)^a}{
\prod_{j=2,4,5,6,7}\Gm(a_j) \Gm(4-a_{4567}-2\ep)(-s)^{a-4+2\ep}}
\nn \\ &&  \hspace*{-48mm}\times
\frac{1}{(2\pi i)^6} \int_{-i\infty}^{+i\infty}
\dd w \prod_{j=1}^5 \dd z_j
\left(\frac{m^2}{-s} \right)^{z_{15}}
\left(\frac{t}{s} \right)^{w}
\nn \\ &&  \hspace*{-48mm}\times
\frac{\Gm(a_2 + w) \Gm(-w) \Gm(z_{24}) \Gm(z_{34})}
{\Gm(a_1 + z_{34}) \Gm(a_3 + z_{24})}
\nn \\ &&  \hspace*{-48mm}\times
\frac{  \Gm(4 - a_{122388}  - 2 \ep + z_{23})
}
{\Gm(4 - a_{46} - 2 a_{57} - 2 \ep - 2 w - z_{1123})}
\nn \\ &&  \hspace*{-48mm}\times
\frac{\Gm(a_{1238} - 2 + \ep + z_{45})\Gm(a_7 + w - z_4)
}
{\Gm(4 - a_{1238} - 2 \ep + w - z_4)}
\nn \\ &&  \hspace*{-48mm}\times
\frac{\Gm(a_{4567} - 2 + \ep + w + z_1 - z_4)}{
\Gm(a_8 - w - z_{234})}
\nn \\ &&  \hspace*{-48mm}\times
\frac{ \Gm(a_8 - z_{234}) \Gm(-w - z_{234} - z_{34})}
{\Gm(4 - a_{13} - 2 a_{28} - 2 \ep + z_{23} - 2 z_5)}
\nn \\ &&  \hspace*{-48mm}\times
 \Gm(2 - a_{567} - \ep - w - z_{12})
\nn \\ &&  \hspace*{-48mm}\times
 \Gm(2 - a_{457} - \ep - w - z_{13})
\nn \\ &&  \hspace*{-48mm}\times
 \Gm(2 - a_{128} - \ep + z_2 - z_5)
\nn \\ &&  \hspace*{-48mm}\times
\Gm(4 - a_{46} - 2 a_{57} - 2 \ep - 2 w - z_{23})
\nn \\ &&  \hspace*{-48mm}\times
\Gm(2 - a_{238} - \ep + z_3 - z_5)
\nn \\ &&  \hspace*{-48mm}\times
\Gm(a_5 + w + z_{234})\Gm(-z_1)\Gm(-z_5)
\, ,
\label{6MB1}
\eea
where $a_{4567}=a_4+a_5+a_6+a_7, a_{122388}=a_1+2a_2+a_3+2a_8$,
$z_{1123}=2z_1+z_2+z_3$, etc.

Analytical evaluation of the master double box
$B_{PL,1}(1,\ldots,1,0)$ was performed in \cite{S3}, with a result
in terms of polylogarithms.
In \cite{BFMR}, planar 2-loop box diagrams  with one-loop insertion
were also by DE.
The general MB representation (\ref{6MB1}) can be used for the
evaluation of other master integrals. As an example,
a double box with a numerator, $B_{PL,1}(1,\ldots,1,-1)$
is evaluated in \cite{HS}.
The finite part of the result in $\ep$ includes polylogarithms and HPL
\[ H\left(-1, 0, 0, 1;-\frac{1-1/\sqrt{1-4m^2/s}}{1+1/\sqrt{1-4m^2/s}}
\right)
\,. \]

\begin {figure} [htbp]
\begin{picture}(170,50)(-30,0)
\Line(-15,0)(0,0)
\Line(-15,50)(0,50)
\Line(115,0)(100,0)
\Line(115,50)(100,50)
\Line(0,0)(50,0)
\DashLine(50,0)(100,0){3}
\DashLine(100,50)(50,50){3}
\Line(50,50)(0,50)
\DashLine(0,50)(0,0){3}
\Line(50,0)(50,50)
\Line(100,0)(100,50)
\Vertex(0,0){1.5}
\Vertex(50,0){1.5}
\Vertex(100,0){1.5}
\Vertex(0,50){1.5}
\Vertex(50,50){1.5}
\Vertex(100,50){1.5}
\end{picture}
\caption{Planar massive on-shell double box of the second type}
\label{2mboxPl2}
\end{figure}
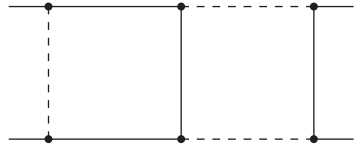

The second planar graph is shown in Fig.~\ref{2mboxPl2}.
For the corresponding general Feynman integral
\bea
B_{PL,2}(a_1,\ldots,a_8;s,t,m^2;\ep)
&& \nn \\ && \hspace*{-48mm}
=\int\int \frac{\dd^dk \, \dd^dl}{(k^2-m^2)^{a_1}(l^2)^{a_6}
[(k-l)^2-m^2]^{a_7}
}
\nn \\ && \hspace*{-48mm}
\times
\frac{1}{[(k+p_1+p_2)^2-m^2]^{a_3}[(l+p_1+p_2)^2)]^{a_4}
}
\nn \\ && \hspace*{-48mm}
\times \frac{[(k+p_1+p_2+p_3)^2]^{-a_8}}{
[(l+p_1+p_2+p_3)^2-m^2]^{a_5}
[(k+p_1)^2)]^{a_2}} \,,
\label{2boxPL2}
\eea
with the same irreducible numerator, the following
sixfold MB representation can be derived:
\bea
B_{PL,2}(a_1,\ldots,a_8;s,t,m^2;\ep)
&& \nn \\ &&  \hspace*{-48mm}
=\frac{\left(i\pi^{d/2} \right)^2 (-1)^a}{
\prod_{j=2,4,5,6,7}\Gm(a_j) \Gm(4-a_{4567}-2\ep)(-s)^{a-4+2\ep}}
\nn \\ &&  \hspace*{-48mm}\times
\frac{1}{(2\pi i)^6} \int_{-i\infty}^{+i\infty}
\prod_{j=1}^6 \dd z_j
\left(\frac{m^2}{-s} \right)^{z_5+z_6}
\left(\frac{t}{s} \right)^{z_1}
\nn \\ &&  \hspace*{-48mm}\times
\prod_{j=1}^6 \Gm(-z_j)
\frac{
\Gm(a_2 + z_1) \Gm(a_4 + z_{24})\Gm(a_6 + z_{34})}
{\Gm(a_3 - z_2)\Gm(a_1 - z_3)
}
\nn \\ &&  \hspace*{-48mm}\times
\frac{ \Gm(4 - a_{445667} - 2 \ep - z_{2344})
}
{\Gm(4 - a_{445667} - 2 \ep -  z_{234455})
}
%\nn \\ &&  \hspace*{-48mm}\times
%\frac{1}{}
\nn \\ &&  \hspace*{-48mm}\times
\frac{
   \Gm(8 - a_{13}-2 a_{245678} - 4 \ep
  - z_{11234455})}
{\Gm(8 - a_{13} - 2 a_{245678} - 4 \ep
  - z_{1123445566})}
\nn \\ &&  \hspace*{-48mm}\times
\frac{\Gm(2 - a_{456} - \ep - z_{45})
\Gm(2 - a_{467} - \ep - z_{2345})
  }
{\Gm(a_{45678}-2 +\ep + z_{2345}) }
\nn \\ &&  \hspace*{-48mm}\times
\frac{\Gm( a_{4567} +\ep -2  + z_{2345})}
{ \Gm(6 -a - 3 \ep - z_{45})}
\nn \\ &&  \hspace*{-48mm}\times
\Gm(a_{45678}-2 +\ep + z_{12345})
\nn \\ &&  \hspace*{-48mm}\times
\Gm(4 - a_{1245678} - 2 \ep -z_{12456})
\nn \\ &&  \hspace*{-48mm}\times
\Gm(4 - a_{2345678} - 2 \ep - z_{13456})
\nn \\ &&  \hspace*{-48mm}\times
\Gm( a -4+ 2 \ep + z_{1456}) \, .
\label{6MB2}
\eea

This multiple MB representation was used in \cite{HS} for the evaluation
of the master planar double box
$B_{PL,2}(1,\ldots,1,0;s,t,m^2;\ep)$.
The result includes polylogarithms, HPL as well
as  two-parametric integrals of elementary functions.
It is not clear at the moment whether these integrals can be
written in terms of
HPL or 2dHPL depending on special combinations
of $s,t$ and $m^2$, or a new class of functions is needed.
This result as well as  many other results for double and triple boxes
mentioned in this short review is confirmed
by numerical evaluation by means of a method
based on a sector decomposition in the space of alpha parameters
\cite{BH}.

\begin {figure} [htbp]
\begin{picture}(170,50)(-30,0)
\Line(-15,0)(0,0)
\Line(-15,50)(0,50)
\Line(115,0)(100,0)
\Line(115,50)(100,50)
\Line(0,0)(50,0)
\Line(50,0)(100,0)
\Line(100,50)(50,50)
\Line(50,50)(0,50)
\DashLine(0,50)(0,0){3}
\DashLine(50,0)(100,50){3}
\DashLine(100,0)(50,50){3}
\Vertex(0,0){1.5}
\Vertex(50,0){1.5}
\Vertex(100,0){1.5}
\Vertex(0,50){1.5}
\Vertex(50,50){1.5}
\Vertex(100,50){1.5}
\Text(25,43)[]{\small 1}
\Text(-7,25)[]{\small 2}
\Text(25,7)[]{\small 3}
\Text(75,7)[]{\small 4}
\Text(60,30)[]{\small 7}
\Text(93,18)[]{\small 5}
\Text(75,43)[]{\small 6}
\end{picture}
\caption{Non-planar massive on-shell double box}
\label{2mboxNP}
\end{figure}
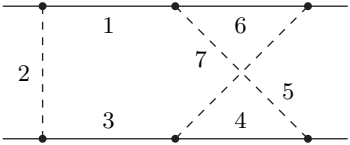

The non-planar graph is shown in Fig.~\ref{2mboxNP}.

The following eightfold MB representation of the general
non-planar double box, with the same irreducible numerator as
above, can be derived:
\bea
B_{NP}(a_1,\ldots,a_8;s,t,u,m^2;\ep)
&& \nn \\ &&  \hspace*{-50.5mm}
= \frac{\left(i\pi^{d/2} \right)^2 (-1)^a}{
\prod_{j=2,4,5,6,7}\Gm(a_j) \Gm(4-a_{4567}-2\ep)(-s)^{a-4+2\ep}}
\nn \\ &&  \hspace*{-50.5mm}\times
\frac{1}{(2\pi i)^8} \int_{-i\infty}^{+i\infty}
\prod_{j=1}^8 \dd z_j
\left(\frac{m^2}{-s} \right)^{z_5+z_6} \!
\left(\frac{t}{s} \right)^{z_7} \! \left(\frac{u}{s} \right)^{z_8}
\nn \\ &&  \hspace*{-50.5mm} \times
\prod_{j=1}^7 \Gm(-z_j)
\frac{\Gm(a_5 + z_{24})
 \Gm(a_7 + z_{34})
 }
{\Gm(a_1 - z_2) \Gm(a_3 - z_3) \Gm(a_8 - z_4)
}
\nn \\ &&  \hspace*{-50.5mm}
\times
\frac{
  \Gm(2 - a_{567} - \ep - z_{245})
  \Gm(2 - a_{457} - \ep - z_{345})
}
{\Gm(4 - a_{455677} - 2 \ep - z_{234455})
}
\nn \\ &&  \hspace*{-50.5mm}
\times
\frac{ \Gm(a_8 + z_{17} - z_4)
\Gm(4 - a_{2345678} - 2 \ep - z_{25678})  }
{\Gm(6 - a - 3 \ep - z_5) }
\nn \\ &&  \hspace*{-50.5mm}
\times
\frac{\Gm(a_{28} + z_{178} - z_4) \Gm(-a_8 - z_{178} + z_4)
}
{\Gm(8 - a_{13} - 2 a_{245678} -4 \ep
- z_{2355667788})}
\nn \\ &&  \hspace*{-50mm}
\times
\frac{\Gm(4 - a_{1245678} - 2 \ep - z_{35678})
}
{\Gm(a_{245678} -2+ \ep + z_{1235788})}
\nn \\ &&  \hspace*{-50mm}
\times
\Gm(a_{4567} + \ep-2 + z_{23458})
\Gm(a-4 + 2 \ep + z_{5678})
\nn \\ &&  \hspace*{-50mm}
\times
\Gm(a_{245678} -2+ \ep +z_{12357788})
\nn \\ &&  \hspace*{-50mm}
\times
\Gm(4 - a_{455677} - 2 \ep - z_{2344})
\nn \\ &&  \hspace*{-50mm}
\times
\Gm(8 - a_{13} - 2 a_{245678} - 4 \ep
- z_{23445588})
    \, .
\label{8MB}
\eea
%It is natural to treat the Mandelstam variables not restricted
%by the physical condition $s+t+u=4m^2$.
This representation can be used for the evaluation of the
non-planar master planar double boxes.
%$B_{NP}(1,\ldots,1,0;s,t,m^2;\ep)$.

\section{Perspectives}

It has been reported \cite{JG} that the reduction of the massive
on-shell double boxes relevant to Bhabha scattering
can be done using Laporta's algorithm. Master integrals for
these three recursion problems were identified. The calculation
of these master integrals was performed in all the cases when
four indices $a_i$ are positive and for some partial cases with
five positive indices. In this problem the method of DE meets
some complications because differential equations
higher that the second order appear for complicated
master integrals.
Whether or not this obstacle can be overcome is an open
question. However, the three general MB representations
presented above can be certainly used for the analytical
evaluation of the master integrals.

Let us finally characterize
advantages of the method based on MB representation:

({\em i}) The MB representation can be derived for a general Feynman
integral corresponding to a
given graph, i.e. with general integer powers of the propagators.

({\em ii}) Resolution of singularities in $\ep$ is much simpler than
in alpha and Feynman parametric integrals.

({\em iii}) After the resolution of singularities in $\ep$ one can always
switch to numerical evaluation, at least in order to check analytical
results. The convergence along imaginary axis is always perfect.
%(it is sufficient to take integration between $-10 i$ and $+10 i$
%instead of $(-i \infty,+i \infty)$ to have accuracy of 10 digits).

({\em iv}) Automation of calculations based on MB representation
looks promising.

\end{document}